\documentclass[utf8]{frontiersSCNS} 
\usepackage{url,hyperref,lineno,microtype,subcaption}
\usepackage[onehalfspacing]{setspace}

\def\mbh{$M_{\rm BH}$\/}

\def\lledd{$L/L_{\rm Edd}$}

\def\rfe{$R_{\rm FeII}$}

\def\feiiq{\rm Fe{\sc ii}$\lambda$4570\/}
\def\msol{M$_\odot$\/}

\def\chm{$c(\frac{1}{2})$\/}
\def\cqm{$c(\frac{1}{4})$\/}
\def\apj{ApJ}
\def\mnras{MNRAS}
\def\ltsima{$\; \buildrel < \over \sim \;$}
\def\ltsim{\lower.5ex\hbox{\ltsima}}  
\def\gtsima{$\; \buildrel > \over \sim \;$}

\def\gtsim{\lower.5ex\hbox{\gtsima}}

\def\civ{{\sc{Civ}}$\lambda$1549\/}

\def\cm3{cm$^{-3}$\/}
\def\hb{{\sc{H}}$\beta$\/}

\def\oiiiopt{{\sc{[Oiii]}}$\lambda\lambda$4959,5007\/}

\def\feiiopt{{Fe \sc{ii}}$_{\rm opt}$\/}

\def\fe{{\sc{Fe}}\/}

\def\fe76087{{\sc [Fe vii]}$\lambda$6087\/}
\def\oiii{{\sc [Oiii]}$\lambda$5007}

\def\kms{km~s$^{-1}$}

\def\rk{$R_{\rm K}$\/}
\def\gs{$\Gamma_{\rm S}$\/}
\def\ergss{erg s$^{-1}$\/}

\def\rk{{$R{\rm _K}$}\/}

\def\lbol{$L_\mathrm{bol}$\/}
\def\nat{\rm Nature\/}
\def\aap{\em  A\&Ap\/}
\def\apss{\em  Ap\&SS\/}
\def\pasj{\em  PASJ\/}

\def\aj{\em AJ\/}
\def\apjs{\em ApJS\/}
\def\apjl{\em ApJL\/}
\def\araa{ARAAp}
\usepackage{enumerate}

\def\keyFont{\fontsize{8}{11}\helveticabold }
\def\firstAuthorLast{Fraix-Burnet {et~al.}} 
\def\Authors{Didier Fraix-Burnet\,$^{1,*}$, Paola Marziani\,$^{2}$ Mauro D'Onofrio\,$^{3}$ and Deborah Dultzin\,$^{4}$}
\def\Address{$^{1}$Univ. Grenoble Alpes, CNRS, IPAG, F-38000 Grenoble, France;\\  $^{2}$INAF, Osservatorio Astronomico di Padova, Italy; \\ $^{3}$Dipartimento di Fisica \& Astronomia, Universit\`a\ di Padova, Italia;\\ $^{4}$Instituto de Astronom\'\i a, UNAM, Mexico
 }
\def\corrAuthor{Didier Fraix-Burnet}
\def\corrEmail{didier.fraix-burnet@univ-grenoble-alpes.fr}
\begin{document}
\onecolumn
\firstpage{1}
\title[Phylogeny and ontogeny of quasars]{ {The} phylogeny of  quasars and  {the} ontogeny of their central black holes} 
\author[\firstAuthorLast ]{\Authors} 
\address{} 
\correspondance{} 
\extraAuth{}
\maketitle

\begin{abstract}

The connection between multifrequency quasar observational  and physical parameters related to accretion processes is still open to debate. In the last 20 year,  Eigenvector 1-based approaches  developed since the early papers by \citet{borosongreen92} and \citet{sulenticetal00b}  have been proven to be a remarkably powerful tool to investigate this issue, and have led to the definition of a quasar ``main sequence''.   In this paper we perform a cladistic analysis on two samples of 215 and 85  low-$z$ quasars ($z \lesssim$ 0.7) which were studied in several previous works and which offer a satisfactory coverage of the Eigenvector 1-derived main sequence. The data encompass accurate measurements of observational parameters which represent key aspects associated with the structural   diversity of quasars. Cladistics  is able to group sources radiating at higher Eddington ratios, as well as to separate  radio-quiet (RQ) and radio-loud (RL) quasars. The analysis suggests a black hole mass threshold for powerful radio emission and also properly distinguishes  core-dominated and lobe-dominated quasars, in accordance with the basic tenet of RL unification schemes. Considering that   black hole mass  provides a sort of ``arrow of time'' of nuclear activity, a phylogenetic interpretation becomes possible if cladistic trees are rooted on black hole mass: the ontogeny of  black holes is represented by their monotonic increase in mass. More massive radio-quiet Population B sources at low-$z$\ become a more evolved counterpart of Population A i.e., wind dominated sources to which  the ``local'' Narrow-Line Seyfert 1s belong.

\tiny
 \keyFont{ \section{Keywords:  black hole physics -- quasars: general -- quasars: evolution -- quasars: radio loud   -- radio jet formation -- multivariate analysis -- cladistics }  } 
\end{abstract}

\section{Introduction}

Quasars represent the  most luminous stable sources in the Universe \citep[e.g., ][]{donofrioetal12,donofrioetal16}, and it is disturbing that some basic issues  still remain  unsolved in spite of decade-long efforts \citep[e.g.,][and references therein]{sulenticetal12a,antonucci13}. The overall most pressing issue is perhaps the inability to connect physical parameters (such as black hole mass; \citealt[][for recent reviews]{marzianisulentic12,shen13,peterson14})  to the observational properties of individual quasars.  This involves a comprehensive  structural and dynamical modeling of the broad line emitting regions \citep[e.g.,][and references therein]{netzer13} which is still missing. In addition, there is no clear explanation of  the existence of a RL and a RQ quasar population  \cite[see e.g.,][for a recent review]{padovani16}.  

During the last two decades it has become clear that one single model cannot fit all type-1 RQ quasars, and { that the diversity of  most observational properties  cannot be associated  only with a viewing angle (i.e., the orientation angle between the accretion disk axis and the line-of-sight)}. The landmark work of  \citet{borosongreen92} identified two main Eigenvectors in a space of parameters defined by radio, optical, UV and X quasars properties of Palomar-Green quasars. The second eigenvector was found to be correlated with luminosity i.e., a reformulation of the well-known Baldwin effect, the anti-correlation between high-ionization line equivalent width and luminosity \citep{baldwinetal78,bianetal12}.  The interpretation of the first eigenvector (hereafter E1), dominated by an anti-correlation between  full-width at half-maximum FWHM(\hb) and \oiiiopt\ and \feiiopt\ prominence,  was however not so clear since the beginning, and has  fostered  speculations on the role of black hole mass,  chemical composition, black hole spin,  Eddington ratio, and orientation \citep[see][for a recent review]{sulenticmarziani15}.  While orientation effects are probably present in RQ quasars \citep{netzeretal92}, they cannot account for the great diversity of observational properties of the line emitting regions, some of them pointing to differences in physical condition, chemical composition and dynamics \citep[e.g.,][]{elvis00,marzianietal01,chelouchenetzer03,nagaoetal06,netzertrakhtenbrot07,shenho14}. 

Fig. 1 shows the main sequence (MS) in  the plane \feiiopt\ prominence (\rfe= I(\feiiq)/I(\hb)) and FWHM of broad \hb\  \citep{borosongreen92,sulenticetal00a,shenho14}.  Several radio, IR, optical, UV, and X properties of low-$z$ quasars (listed in Table 1) can be organized along the MS.  To focus on the main physical aspects, \citet{sulenticetal00b} defined a 4D parameter spaces with parameters that are observationally orthogonal: in addition to \rfe\ and FWHM(\hb) which are indicators of the low-ionization emitting gas physical conditions and dynamics, they considered the high-ionization line \civ\ shift with respect to the systemic reference of quasars (a tracer of powerful outflows, \citealt{richardsetal11}), and the soft X-ray photon index \gs, related to the accretion mode \citep[e.g.,][]{poundsetal95,grupe04}.  

The data in Table 1 are reported   for typical Population A   (FWHM $\lesssim$\ 4000 \kms) and Population B (broader) sources following the  definition of \citet{sulenticetal00a}. The separation into two populations  was originally proposed on the basis of a rather abrupt discontinuity in the shape of the Balmer line profiles \citep[e.g.,][]{sulenticetal02,collinetal06}. Systematic changes at FWHM $\sim$\ 4000 \kms\ are probably associated with a discontinuity in accretion mode \citep[][]{marzianietal03b,marzianietal14}, as the most relevant physical parameter governing the MS and the differences summarized in Table 1 is likely   Eddington ratio  \citep{marzianietal01,shenho14}. 

Observational differences for sources belonging to the  {two} Populations become obvious looking  at the extremes of the MS.  At the top left positions on the sequence shown in Fig. \ref{fig:e1} (Population B), lines are very broad (often with composite and couple peaked profiles, \citealt{stratevaetal03}), redward asymmetric, \feiiopt\ is weak, the spectral energy distribution (SED) is hard \citep{laoretal97b,shangetal07}, \oiiiopt\ is strong and symmetric with a blueward asymmetry close to the line base.  At the other end,  extreme Pop. A sources (\rfe $\gtrsim$ 1) are high accretors which show evidence of strong radiation driven winds \citep{sulenticetal07,richardsetal11,coatmanetal16}. 

All these developments in the study of quasars strongly indicate that statistical multivariate analyses could be useful. They should bring a more powerful capacity to determine the relevant properties that explain the observed diversity. In this paper, since quasars are known to be evolving objects, we have chosen to use a phylogenetic approach which establishes evolutionary relationships. Among the possible tools, cladistics is certainly the most general and simplest to implement. 

In the following, we first describe the cladistic method (\S\ \ref{method}). The method is then applied to a set of optical and UV parameters that include the four parameters of the \citet{sulenticetal00b} 4D space, as well as several additional key parameters  (\S\ \ref{param}). They are available for a sample of 85 sources (Sect. \ref{sample}) which is as subsample of 215 sources for which only the optical spectral range is available and that is also considered in this study. The cladistic analysis resolves the difference between Pop. A and B and indicates that RL sources are the most evolved Population B sources   (\S\ \ref{results}). The results are briefly  discusses in terms of quasar populations at low-$z$ as well as of evolution over cosmic age (\S \ref{onto}), and in light of our present understanding of the RQ/RL dichotomy (\S \ref{rqrl}).

\section{Method}

\label{method}

\subsection{Cladistic Analysis}

Astrocladistics aims at introducing phylogenetic tools in astrophysics. A phylogeny shows the evolutionary relationships between groups, species or classes, while a genealogy shows the relationships between individuals with parents and offsprings. Ontogeny is the evolution of a single individual. In astrophysics, the observations only allow for phylogenetic reconstructions. Consequently, in this paper, each quasar supposedly represents a species (or sub-species, class...). Nevertheless, for simple objects, like a black hole or a star, it is possible to observe identical objects at different stages of evolution, so that their ontogeny can be derived.  Among the tools developed for phylogenetic analyses, cladistics, also called Maximum Parsimony, is the most general and the simplest to implement. It uses parameters, and not distances, to establish the relationships between the species by minimizing the total evolutionary cost depicted on a phylogenetic tree.

\subsubsection{Outline of the astrocladistics approach}

The cladistic analysis is a phylogenetic method for classification, an unsupervised multivariate classification technique that establishes the relationships between the taxa under study. A taxon is a class, a species, or an individual supposedly representing a class. The relationships are depicted on a phylogenetic tree that represents the simplest evolutionary scenario given the data. 

Contrarily to many clustering and phylogenetic techniques, cladistics does not compute distances between the taxa, but uses the parameters themselves. These parameters ideally must bear an information of evolution, so that they can be discretized into evolutionary stages. In this case they are called "characters", and the number of changes in stages (steps) represents an evolutionary cost. For a given tree, the total number of steps, considering all the characters and all the taxa, defines the complexity of the evolutionary scenario depicted by the tree. The cladistic algorithm aims at finding the simplest evolutionary scenario which is given by the most parsimonious tree, the one that has the lowest total number of steps, among all possible tree topologies that can be constructed with the taxa in the sample. This kind of tree is called a cladogram. Because it is not based on distances, the cladistic algorithm accepts undocumented values.

The reader is referred to \citet{Fraix-Burnet2015} for a review on unsupervised classification techniques in extragalactic astronomy, and to \citet{jc1,jc2,Fraix2012,Fraix-BurnetHouches2016} for more details on the use of cladistics in astrophysics. {More explanations are available at https://astrocladistics.org.}

\subsubsection{Implementation of the cladistic analyses}

As usual in astrocladistics, we discretized each parameter in each sample into 32 bins in order to keep its continuous and quantative nature. The optimization criterion that we use considers that the evolutionary cost between two states is the absolute value of their difference (l1-norm). The computation used the heuristic search algorithm implemented in Phylogenetic Analysis Using Parsimony PAUP*4.0b10 \citep{paup}, with a ratchet method to avoid as much as possible local minima \citep{ratchet}.

\subsubsection{Reliability assessment}

The reliability (``robustness") of the cladistic analysis has been estimated through several complementary analyses taking slightly different subsets of quasars and parameters. The two samples presented in this paper are the best ones for two reasons. Firstly, the consensus tree obtained from all the most parsimonious trees found for each sample is nearly entirely resolved, showing that they are all in excellent agreement. This indicates that the tree structure corresponds very probably to a global minimum. Secondly, the results with these two samples agree  {very well with the complementary analyses}, some slight disagreement occurring at the relative placement of some groups on the trees. This does not affect the main interpretation of the proposed phylogeny of quasars.

\subsubsection{Interpretation of the tree}

The phylogenetic tree depicts the shortest path to relate all the objects of the sample through an evolutionary history. Starting from any taxon (a leave on the tree), one can estimate the cost to transform this object into any other one. 

The tree is a hierarchical organization of the taxa, and there is no objective way to define species or classes. The substructures of the tree are a good indication that the related objects may be close from an evolutionary point of view. Ideally, a clade is composed of all the taxa that are the descendants emerging from an internal node (i.e. an unlabelled node). These taxa are supposed to have inherited a property from a common (unknown) ancestor placed at the internal node. Hence each substructure has some chance to correspond to a clade and are used here to define the evolutionary groups of quasars.

The detailed structure of the tree depends somewhat on whether the sense of evolution has been imposed. This is made by defining a root, i.e. a taxon or a clade,  which is the closest to the ancestor common to all the object of the sample. This is of course not an easy choice to make in a multivariate pattern. One may use a few parameters that are known to evolve in a monotonic fashion, such as the mass of the black hole which cannot decrease easily if at all. One must be careful that this approach does not lead to a conflict with other parameters.

In any case, once the tree is built and can be trusted statistically, it should be seen as a phylogenetic hypothesis extracted from the data, an hypothesis to be confronted with the current astrophysical knowledge.

\subsection{Sample selection} 
\label{sample}

The selected samples  have the non-negligible advantages to cover relatively well  the spectral bins of largest occupation along the E1 MS (Fig. \ref{fig:e1}), i.e., to provide a fair representation of the quasar spectroscopic diversity at low-$z$. The evolution of the quasar luminosity function derived by \citet{boyleetal00} for $z \lesssim 0.7$\ is modest.  Quasars with \rfe $\lesssim$ 1.5  are 98 \%\ of all low-$z$ quasars \citep{marzianietal13a}.  It is worth noting that both samples are strongly biased in favor of RL quasars. This is not really a hindrance since an   optically-selected, flux-limited complete sample of 85 sources should include only 6 -- 7 RL sources, encompassing both core-dominated (CD) and lobe-dominated (LD) quasars.    The main results related to the MS have been confirmed by the eventual analysis of large Sloan Digital Sky Survey (SDSS)-based samples  \citep[e.g.,][]{zamfiretal10,shenho14}.  

The larger sample (M215) includes 215 low-$z$\ quasars ($z \lesssim 0.7$) presented by \citet{marzianietal03a}. Measurements of \oiii\ are available for most of these sources in   addition to FWHM(\hb) and \rfe\ but the sample lacks UV and soft-X ray information.  A second sample (M85) includes  85 sources, and is obtained  from the intersection of the \citet[][215 sources covering the \hb\ spectral range]{marzianietal03a} and the \citet[][130 sources with  \civ\ covered from Hubble Space Telescope Faint Object Spectrograph  (HST/FOS) observations]{sulenticetal07} samples. The measurements  used in this paper are available on Vizier for both {\citet{marzianietal03a}\footnote{\tt http://cdsarc.u-strasbg.fr/viz-bin/Cat?J/ApJS/145/199} and \citet{sulenticetal07}\footnote{{\tt http://cdsarc.u-strasbg.fr/viz-bin/Cat?J/ApJ/666/757}}. The two samples provide reliable measurements on high S/N, moderate dispersion spectra that are unmatched by the automated measurements obtained on the SDSS spectra by other authors. M85 is necessary for characterizing the dynamics of the line emitting regions, as there is strong evidence that a virialized component (traced by \hb) is coexisting with an outflowing component traced by \civ \citep[e.g.,][Sulentic et al. 2017, in preparation]{marzianietal16}. The black hole mass (\mbh) has been estimated from \hb\ FWHM using the \citet{vestergaardpeterson06} scaling law, and the Eddington ratio computed assuming a factor 10 bolometric correction for the measured specific flux at 5100 \AA\ i.e., applying a correction consistent with past and more recent estimates \citep{elvisetal94,richardsetal06}.

 \begin{figure}[h!]
\begin{center}
\includegraphics[width=10cm]{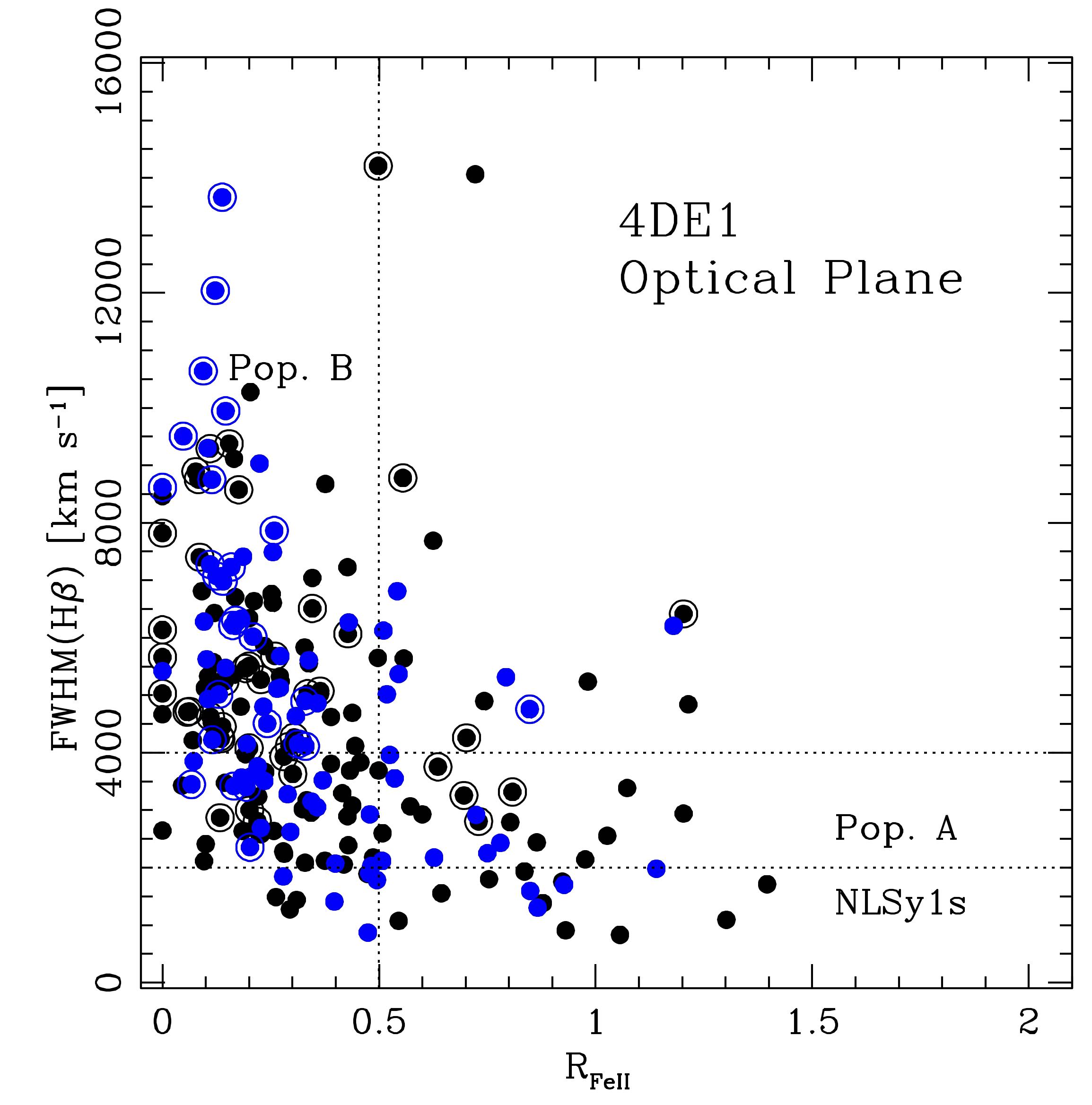}
\end{center}
\caption{The optical plane of the 4DE1, FWHM(\hb) vs. \rfe\ for the M215 sample. Sources belonging  to   M85 are in blue, and radio loud sources are represented by circled symbols. The limit of Pop. A and B are marked, as well as the conventional limit of narrow-line Seyfert 1s (NLSy1).  } \label{fig:e1}
\end{figure}

\subsection{Parameter selection {for the cladistic analysis}}
\label{param}

{In a multivariate analysis such as classification, great care should be taken to avoid redundancies or uninformative parameters that could perturb the result. As already said, the characters are expected to trace out the evolution of the quasars, and obviously, correlated parameters may put too much weight on the underlying physical process. However some correlations are not causal \citep{DFB2011} and thus should not be eliminated. The disturbing parameters bring noise and can prevent the convergence of the analysis. It is thus important to understand well the parameters for the sample under study and then repeat the cladistics analysis with several subsets of parameters to test the robustness of the phylogenetic signal we are looking for. For the interpretation of the tree, naturally, all available information can be used and any parameters can be projected onto the tree.}

Among the available parameters here, the absolute $B$ magnitude\ $M_\mathrm{B}$\ and \lbol\ are both measures of luminosity and consequently are strongly correlated. We keep only \lbol\ for the analysis. The black hole mass  \mbh\ is   a derived property and it is included in the results presented in this paper, but analyses performed without this parameter yields very similar results with a slightly less resolved tree. The Eddington ratio is also a derived quantity and was not included in the cladistic analysis presented in this paper. If we include the Eddington ratio in addition to \mbh, then the trees become more linear, pointing to a correlation between two or more parameters.  As a consequence, we prefer not to use it to establish the tree, and include it for the interpretation only.

The M215 sample finally includes seven parameters  {for the cladistic analysis}: radio loudness parameter \rk\ (coded as {\tt logRK} in the Figures) following \citet{kellermannetal89}, \rfe\ ({\tt RFE}), FWHM(\hb) ({\tt FWHMHb}), the line centroid displacement of \hb\ at quarter maximum \cqm\  ({\tt c1o4Hb}), W(\oiii) ({\tt WOIII}, in \AA), the peak shift of \oiii\ ({\tt voIII}), the decimal logarithms of  bolometric luminosity \lbol\ ({\tt logLbol}),  \mbh\ ({\tt logMBH}). All radial velocity displacements are referred to the quasar rest frame, in units of \kms.
Note that {\tt voIII} has 28 unknown values.

The M85 sample includes eleven parameters, that is four additional ones with respect to M215: the photon index  \gs\ ({\tt Gamma}), the radial velocity centroid displacement of \civ\  at half maximum \chm\ ({\tt c1o2CIV}), and the  rest-frame equivalent width of \civ, W(\civ)\ ({\tt WCIV}).  There are twelve unknown values for {\tt voIII} and four for {\tt Gamma}.

\section{Results}
\label{results}

\begin{figure}[h!]
\begin{center}
\includegraphics[width=\linewidth]{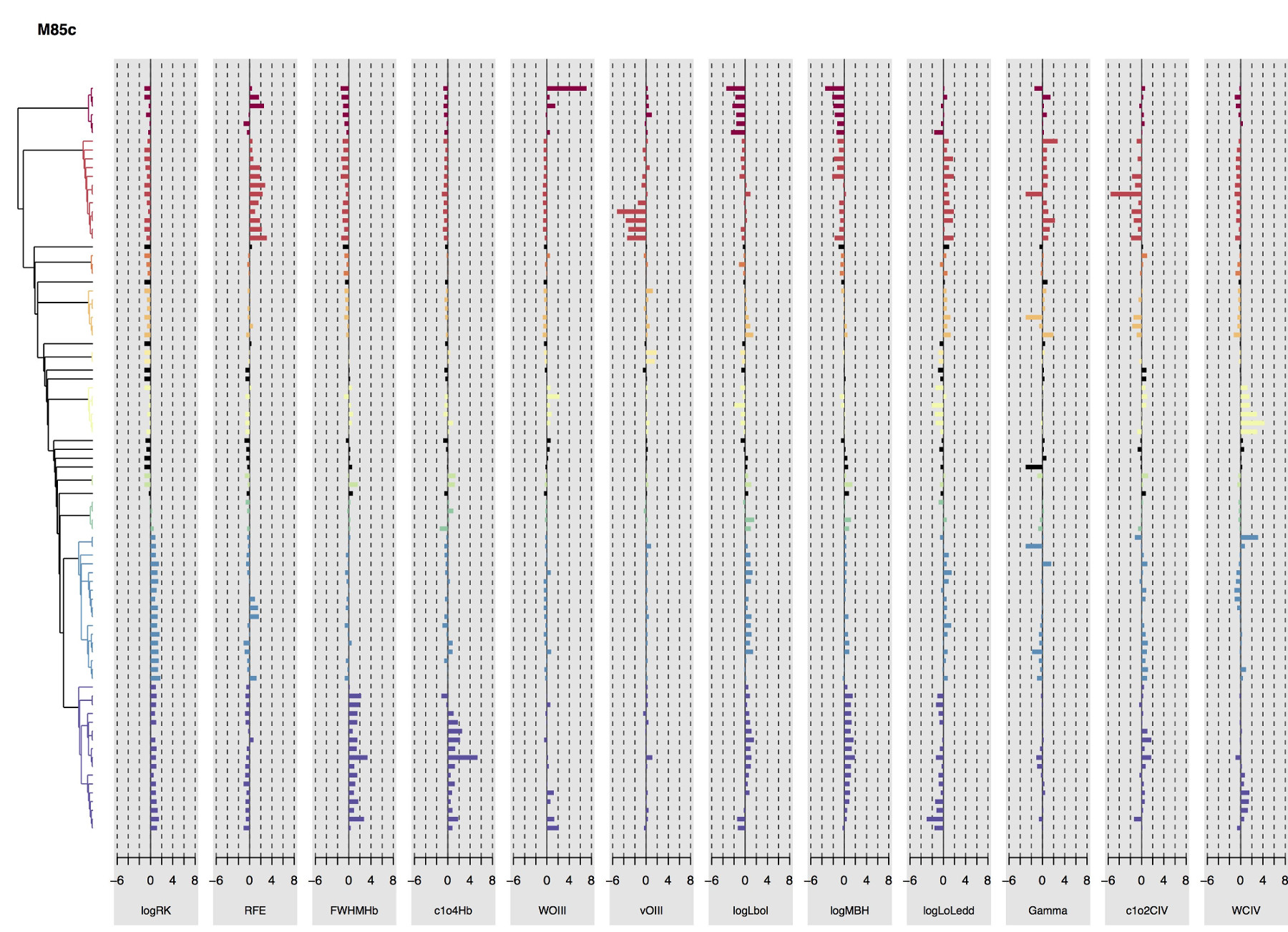}
\end{center}
\caption{Cladograms for the M85 sample rooted with the lowest-\mbh\ group, {showing barplots for the Eddington ratio \lledd\ in addition to the eleven parameters used to find the tree}. Colors correspond to the groups defined from the tree on the left.}\label{fig:clado85}
\end{figure}

\begin{figure}[h!]
\begin{center}
\includegraphics[width=\linewidth]{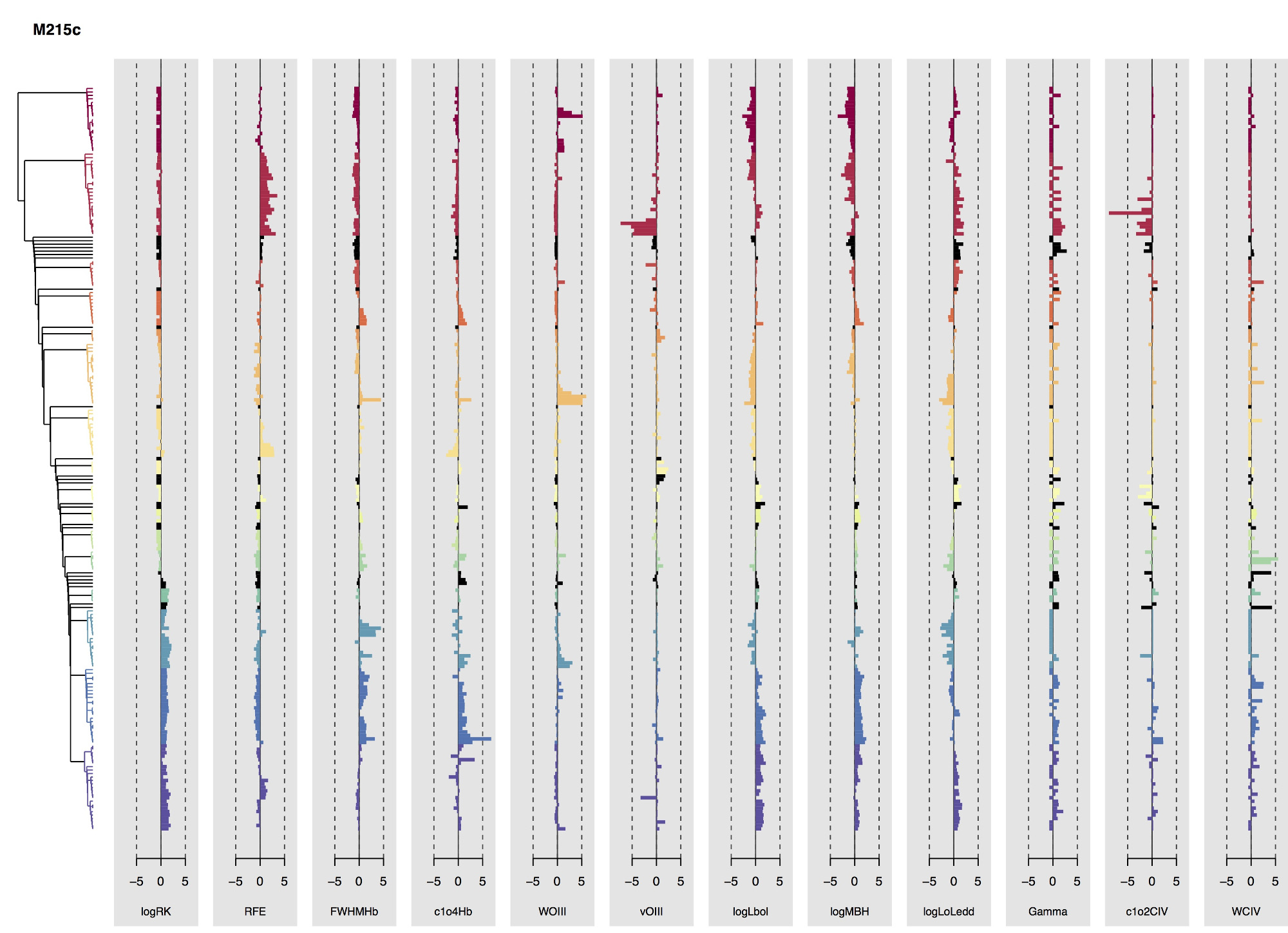}
\end{center}
\caption{Cladogram for the M215 sample rooted with the lowest-\mbh\ group, {showing barplots for \lledd, \gs, \civ, \chm, W(\civ), in addition to the seven parameters used to find the tree}. Colors correspond to the groups defined from the tree on the left. They do not match in any way the groups and colors for the M85 sample, except for the progression from red to deep blue}\label{fig:clado215}
\end{figure}

 \begin{figure}[htp!]
\begin{center}
\includegraphics[width=\linewidth]{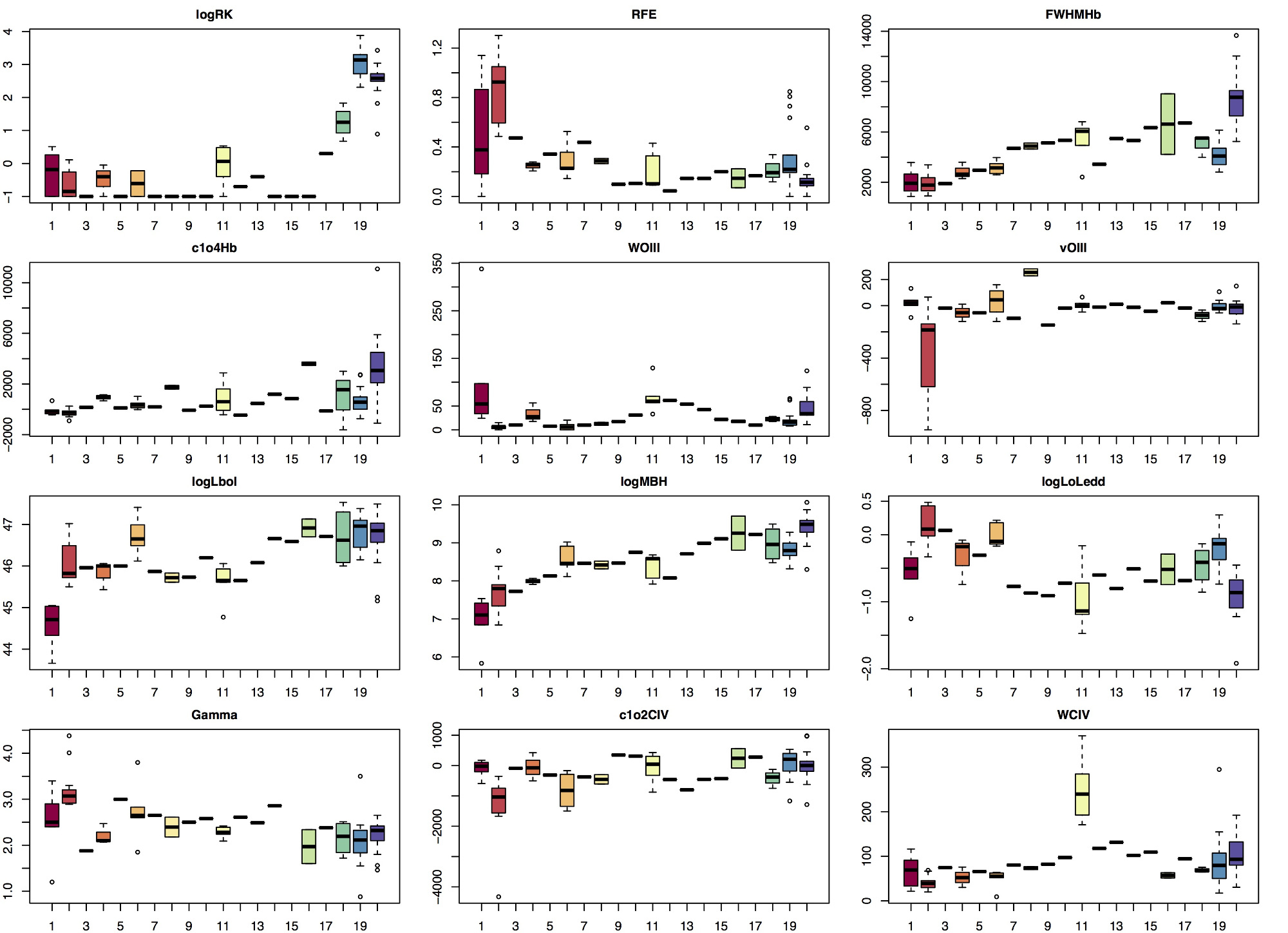}
\end{center}
\caption{ Box plots with identified groups for the M85 sample. The ordinate axis shows the parameter coded on top of the plot, with code meaning  explained in \S \ref{param}. Medians are thick horizontal lines while boxes define limits from first to third interquartile range. Outliers are shown as open circles. Colors correspond  to those in Fig. \ref{fig:clado85}.   }\label{fig:box85}
\end{figure}

 \begin{figure}[htp!]
\begin{center}
\includegraphics[width=\linewidth]{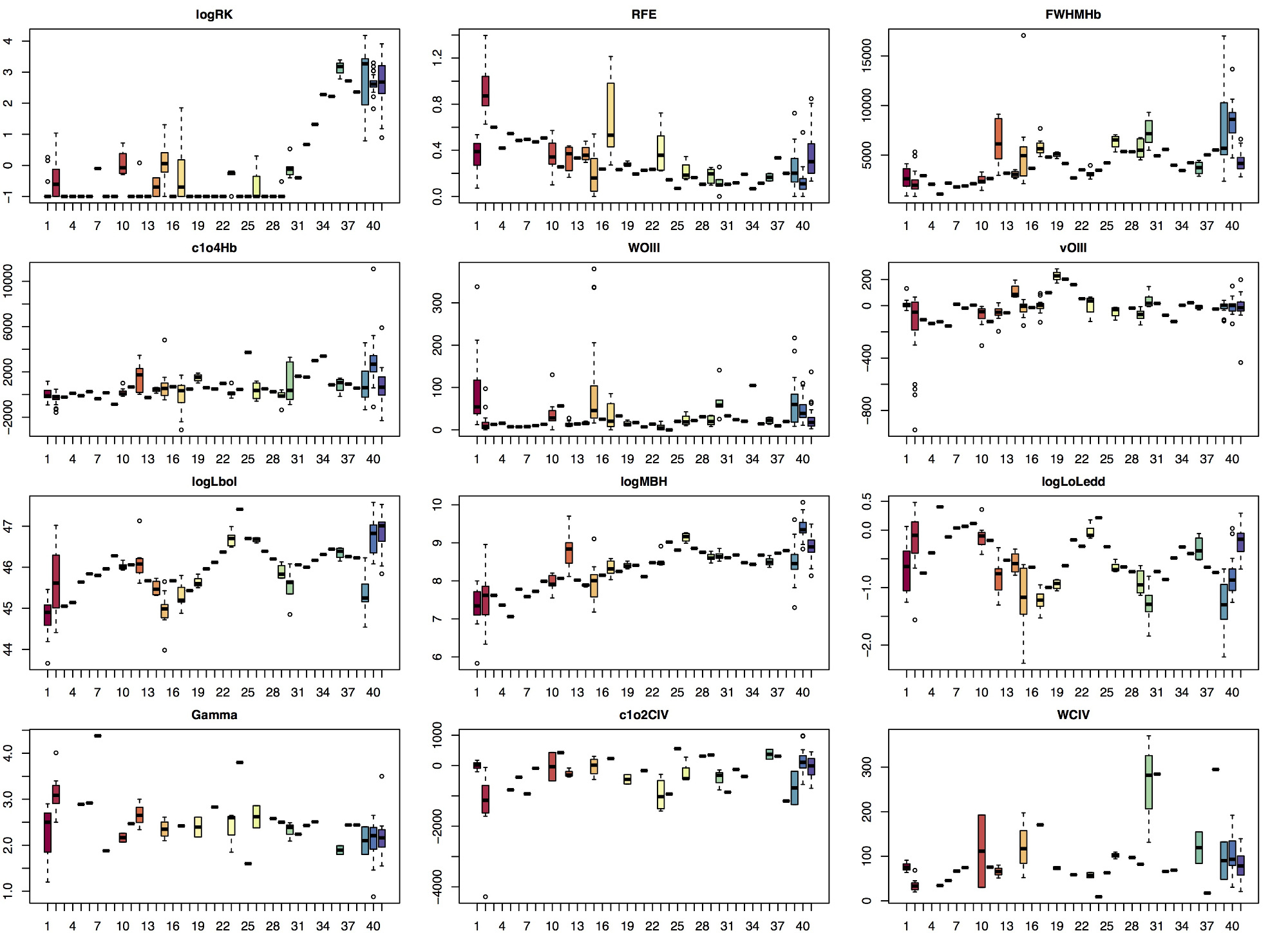}
\end{center}
\caption{ Box plots with identified groups for the   M215 sample. Colors correspond  to those in Fig. \ref{fig:clado215}, and the ordinate axis meaning is the same of Fig. \ref{fig:box85}.  }\label{fig:box215}
\end{figure}

The result of a cladistic analysis is an unrooted tree. It is however convenient to root it for the definition of groups and for a possible evolutionary interpretation. In the case of quasars, a parameter that can root the cladistic trees is black hole mass, since \mbh\ can only grow as a function of cosmic time: the only way a black hole can disappear is through emission of Hawking's radiation  \citep{hawking74} which is tremendously inefficient for massive black holes.  A tree rooted in this way shows a  clustering consistent with evolution from less massive to more massive sources. This evolution would represent the real diversification arrow of quasars if and only if the black hole mass is a reliable evolutionary clock. This would in particular imply that the common ancestor of all quasars has a small black hole. We do not make this assumption in this paper since this parameter is merely one part of the numerous and complex transformation processes of these objects. With this caution stated, the quasar sample contains a population of massive quasars which can be seen as ``more evolved'' than a population of less-massive quasars that are radiating at a higher \lledd. 

The groups can be defined following the substructures of the tree. Formally, clades are (monophyletic) groups of taxa including an hypothetical ancestor (an internal node) and all its descendants. There is no constraint on the level of granularity of the classification, except for homogeneity of the groups and convenience for the interpretation. 

We have chosen the groups as shown in Figs.~\ref{fig:clado85} and ~\ref{fig:clado215}. We find 20 groups on the M85 tree and 41 on the M215 one.\footnote{The group identification is reported for each quasar in the  file {\tt quasargroups.csv} appended to the paper, Table 2.} The relative homogeneity of the groups can be estimated from the barplots drawn in front of the trees, as well as some evolutionary trends along this phylogeny. In addition, there are individual terminal branches (or leaves, i.e. leading to only one quasar) that are identified with black color in this paper. Each can be considered as representative of a group, increasing the total number of identified groups. Note that the color progressions are identical in the two cases. The agreement between the two classifications can be estimated by comparing them with the M85 sample objects only, {with the Adjusted Rand Index that measures the number of pairs of objects that belong or not to the same class in the two classifications. In a perfect match it should be 1, and 0 for random classifications. Here we find 0.42, that} is quite good since the two classifications do not use the same set of parameters. This can also be verified on the boxplots.

Each group can be characterized by the statistics of the parameters and shown as boxplots (Fig.~\ref{fig:box85} and \ref{fig:box215}). From this, we can already identify known populations (Fig.~\ref{fig:tree}): {in M85, group 1 is due to low-luminosity Pop. A sources, group  2 is dominated by extreme Pop. A}, while groups 19 and 20 include almost exclusively radio-loud objects: mainly LDs (20) and CDs (19). Group 18 includes one core-bright LDs  and three radio intermediate sources, $0 \lesssim \log $ \rk\ $\lesssim 1.8$.  Radio-loud groups are from 39 to 41 in M215 (41 predominantly  made of CDs, 40 of LDs). The intermediate groups include low-right corner Pop. A sources but are mainly Pop. B. The cladograms (Figs. \ref{fig:clado85} and \ref{fig:clado215}) trace the relations among the  groups. Even if no evolutionary inference should be inferred {at this point}, it is interesting to note that there is a sequence   going from extreme Pop. A and Pop. B RQ, to Pop. B RL. The bottom groups are core-dominated and lobe-dominated RL sources, which are monophyletic groups in M85.  In M85 and M215, the different groups appear to follow defined systematic trends, with  FWHM(\hb), \rfe\ and \gs\ decreasing from the first group to the last. Large \civ\ blueshifts seem to be restricted to the first group. The cladistic analysis therefore recovers the trends that define the quasars MS in 4DE1 parameter space involving FWHM(\hb), \rfe, \chm\ \civ, and \gs.  High values of the \lledd\ are restricted to the first groups with the smallest masses. Largest \rk\  are associated with the largest \mbh, and LD and CD are separated into two distinct groups (here group \#\ 19 and 20 for M85).  

 At the same time \mbh\ increases quite monotonically, this regularity being independent of the choice to root the tree with the lowest \mbh. {The trend of \mbh\ along the trees is remarkable: there is a strong increase first and then a kind of plateau that may even be split from the first part. This behavior seems strongly related to the trend in \rk, \lledd\ and FWHM(\hb).} {The anomalous group 11 in M85 has large \rk, \large W(\civ): most likely the reflection of large extinction internal to the quasars belonging to this small group. Apart from this, it is apparently not different from the rest of Population B groups.}

\subsection{A threshold \mbh\ for RLness}
\label{threshold}

RL quasars are predominantly found among Pop. B, where RQ quasars are also found with similar accretion parameters {($45 \lesssim \log$ \lbol\  $\lesssim$ 46 \ergss, $8.5 \lesssim \log$ \ \mbh\ $\lesssim$ 9.5, \lledd $\sim$ 0.1)}.  Powerful RL sources appear in our low-$z$ sample only for large \mbh, and CD  and LD are separated because of an intervening role of orientation (which affects FWHM(\hb), and \lbol; in phylogenetic terms, CD and LD belong to different monophyletic groups). We insist on the fact that the other analyses that we have performed without \mbh\ give the same results, so that this conclusion does not depend on the presence of this parameter in the analysis nor on its selection for the rooting of the tree. The \mbh\ result is clearly dependent on sample properties. M215 and M85 contains only very powerful radio-loud sources, so that results from additional studies need to be considered to decide whether a mass threshold is appropriate (\S \ref{rqrl}).

The most established tenet of radio-loud unification \citep{urrypadovani95} is that the difference between CDs and LDs is due to relativistic beaming and therefore strongly dependent on orientation. It is also known that CD and LD sources show differences in optical spectroscopic properties \citep[e.g.,s][]{willsbrowne86,zamfiretal08,buttiglioneetal10}.  We do not really see a separation that may be ascribed to orientation effects in the other branches and trunk of the cladistic tree. Therefore orientation appears to be an intervening physical parameter that differentiates the RL sources into CDs and LD staring from a common progenitor species.

 \begin{figure}[ht]
\begin{center}
\includegraphics[width=16cm]{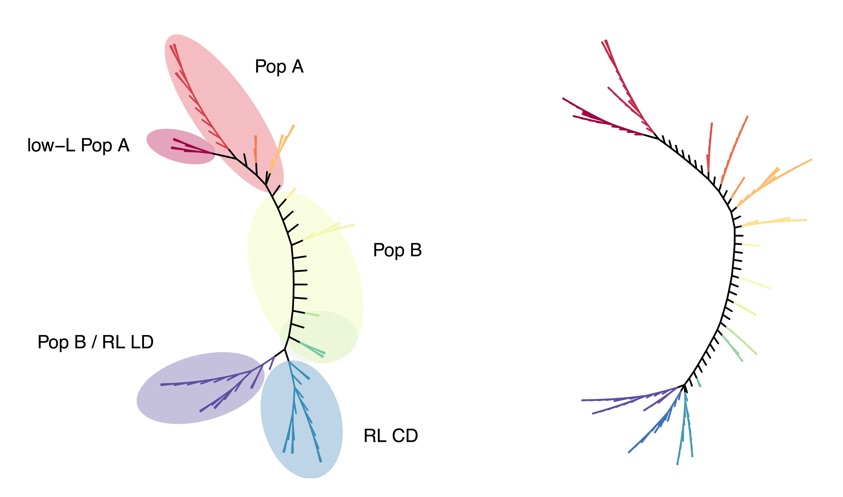}
\end{center}
\caption{Unrooted cladistic trees for the M85 (left) and M215 (right) samples. The main groups are identified by shaded areas in the M85 diagram. Colors corresponds to those in Fig. \ref{fig:clado85} (M85) and Fig. \ref{fig:clado215} (M215). }\label{fig:tree}
\end{figure}

\section{Discussion}

\subsection{\lledd}
Several observational parameters are strongly correlated with \lledd\ \citep{marzianietal01,kuraszkiewiczetal04,shenho14}. 
\lledd\  expresses the ratio between radiation and gravitational forces \citep{netzermarziani10,marzianietal10}.  The transition between Pop. A and B   occurs at  \lledd $\approx 0.2 \pm 0.1$\ \citep{marzianietal03b}, which is consistent with the limit that corresponds to the transition from a geometrically thin to a geometrically thick accretion flow sustained by radiation pressure \citep[][and references therein]{abramowiczstaub14}. Prominent winds are associated with high \lledd\ \citep{sulenticetal07,richardsetal11,coatmanetal16}.    Therefore, we can think of two quasars populations, one with relatively modest \mbh \ ($7 \lesssim \log$ \mbh $\lesssim$ 8 [\msol])  radiating at high Eddington ratio and associated with strong wind, and one of more massive sources ($8 \lesssim \log$ \mbh $\lesssim$ 10 [\msol])  radiating at \lledd $\lesssim 0.1$.   While \lledd\ appears to be the main physical factor governing E1, high-\mbh\ quasars may have resembled low-\mbh\ quasars in an earlier stage of their evolution when they were ``wind-dominated'' (as further discussed in \S \ref{onto}).

\subsection{Ontogenesis and phylogenesis of quasars}
\label{onto}

\subsubsection{Population B as evolved Population A quasars}

Every quasar is the direct descendant of a seed black hole. However, black holes of masses \mbh$\lesssim 10^{5}$ \msol\ are extremely difficult to detect if they are placed in the nuclei of external galaxies. Even if they are radiating at \lledd $\sim 1$ \citep[\lledd\ $\gtrsim$ 2 --  3 may not be possible][]{mineshigeetal00}, their apparent $V$ magnitude will be $\approx$ 22 at redshift $z \approx 0.3$. In flux-limited quasar  samples at low-$z$, we detect  quasars in the mass range 6 $\lesssim \log $ \mbh $\lesssim$ 8 radiating close to their Eddington limit. The masses of these sources are clearly not the masses of the fledgling seed BHs. Nonetheless, in the local Universe, the only sources radiating close to the Eddington limit are these relatively low-\mbh\ quasars. It is easy to see  that, if the most massive black holes were nowadays radiating at their maximum radiative power per unit mass ($\approx 2$ \lledd), they would be almost visible to the naked eye (with \mbh $\approx 10^{10}$\ \msol\ at $z \approx 0.15$\ it would be $m_\mathrm{V} \approx 6.7$!). The absence of massive BHs radiating close to the Eddington limit  is associated with the overall  downsizing of the star formation and nuclear activity at recent cosmic epochs \citep[e.g., ][]{fontanotetal09,revigliohelfand09,hirschmannetal14}. The very massive black holes that were shining bright  mostly  belong now to spent systems \citep[][]{lynden-bell69}, accreting at a very low rate. In this respect sources like Messier 87 (which we consider a prototypical example of the ``spent'' quasars which is hosting one of the most massive BHs in the local Universe, \citealt{walshetal13}) are very different from the Pop. B sources that we are considering in our sample. Pop. B sources are quasars accreting at a modest pace but high enough to be in a radiatively efficient accretion mode that can be   modeled by a geometrically thin, optically thick $\alpha$-disk with an efficiency $\eta \sim 0.07$\ \citep[][]{shakurasunyaev73}. Assuming that Pop. B sources are accreting at a constant mass rate, \mbh\ grows in a linear regime,  the time needed for a Pop. B source to have grown from a mass $M_\mathrm{BH}^{0}$ = $10^{8}$ \msol\ is 
\begin{equation}
\Delta t =  \frac{c^{2}(M_\mathrm{BH} - M_\mathrm{BH}^{0})}{L} \frac{\eta}{1-\eta} \approx 5.7 \cdot 10^{7} M^{0}_\mathrm{BH, 8} (f -1) L^{-1}_{46} \eta_{0.1}\ {\rm yr}
\end{equation}
where $f$\ is the ratio of the actual mass to the initial one and we have assumed $\eta \ll$ 1 \citep[][p. 297]{netzer13}. The $\Delta t$\ needed for a BH to grow from the typical masses of the highly accreting quasars  at a typical luminosity of luminous low-$z$ quasars   is short relative to the expectation of the cosmic evolution of quasar accretion rates. For example, a Pop. A quasar with   $M_\mathrm{BH}^{0}\approx 10^{8}$ \msol\ at $z\approx 0.2$\ could  be seen as a Pop. B source with \mbh $\approx 10^{9}$ \msol\  at $z\approx 0.15$\ since $\Delta t \approx 5.1 \cdot 10^8$\ yr.  Its spectroscopic appearance would change as well, moving the source from the bottom right toward the top left of the E1 MS, as schematically indicated in Fig. \ref{fig:e1int}. In other words, it is legitimate to assume that Pop. B  are evolved Pop. A sources and not the same quasars that were once radiating at very high  $L$\ at the cosmic peak of star formation rate and quasar activity.

This eventuality is rather unlikely also if we consider the expected comoving density of spent (dead) quasars in the local Universe.  It is convenient to follow the \citet{boyleetal00} luminosity function parameterization: 

\begin{equation}
\Phi (M_\mathrm{B},z) = \frac{\Phi(M_\mathrm{B}^{*})}{10^{0.4[(a+1)(M_\mathrm{B}-
M_\mathrm{B}^{*}(z))]} + 10^{0.4[(b+1)(M_\mathrm{B}-M_\mathrm{B}^{*}(z))]}},
\end{equation}

where the evolution is given by the redshift dependence of the break magnitude $M_\mathrm{B}^{*}(z)  =M_\mathrm{B}^{*}(0) - 2.5 (k_1 z + k_2z^2)$, and $a \approx -3.41$, $b \approx  -1.58$. The most luminous quasars are the ones with the most massive black holes radiating close to the Eddington limit (which were relatively rare at $z \approx$ 2 and have all but disappeared at $z \lesssim 0.7$. It is reasonable to assume that all extremely luminous quasars have \mbh  $\lesssim 10^{9.5}$ \msol\ \citep[][]{sulenticetal06,natarajantreister09,king16}, and that they may not radiate at highly super-Eddington ratios, the limit being close to a few times the Eddington ratio. For a typical quasar SED, $\log L \approx 36.54 - 0.4 M_\mathrm{B}$. Therefore, with $L_\mathrm{Edd} \approx 1.3 \cdot 10^{5} \frac{M_\mathrm{BH}}{M_{\odot}}$ $L_{\odot}$, 
$ M_\mathrm{B,max} \approx -27.9$\ for \mbh $ = 10^{9}$ \msol\ and \lledd = 1. The comoving number density $\tilde{n}$\ of sources  with $M_\mathrm{B}$  below this absolute magnitude limit yields a number of sources $dN$\ in the comoving volume element { $dV$t}, and   the number of sources  $N_\mathrm{lum}$\ which are supermassive (\mbh\ $\gtrsim 10^{9}$ \msol) and that were once radiating close to their Eddington limit can be obtained by integrating over $1.5 \lesssim z \lesssim 2.3$:

\begin{equation}
N_\mathrm{lum} = \int_{z_{1}}^{z_{2}}\tilde{n} dV  = \int_{z_{1}}^{z_{2}} \left( \int^{M_\mathrm{B,max}}_{-\infty} \Phi (M_\mathrm{B},z) dM_\mathrm{B}  \right) 4 \pi d_\mathrm{C}^{2}\frac{dz}{E(z)},
\end{equation}
where $d_\mathrm{C}$ is the comoving distance, and $E(z) = \sqrt{\Omega_\mathrm{M}(1+z)^{3} + \Omega_{\Lambda}}$. 
This number can be compared to the number $N_\mathrm{B}$\ of Pop. B sources  up to $z \approx$ 0.75    i.e., to $N_\mathrm{B} = \int_{z_{1}}^{z_{2}}\tilde{n} dV$, with  $z_{1} \approx 0.1$ and $z_{2} \approx 0.75$.   {Assuming that Pop. B are   about $\frac{1}{2}$\  of all type-1 quasars of all quasars within $   M_\mathrm{B,max} \  \le -21$  (as suggested by flux-limited optically selected samples \citealt{zamfiretal10}),  the  number of very massive black holes that were once radiating above --27.9 in the redshift range $1.5 \le z \le 2.3$  falls short by slightly less than two orders of magnitude to explain the density of the ``present-day'' ($z \lesssim$ 0.75) Population B quasars:  $N_\mathrm{lum} \approx 0.03 N_\mathrm{B}$.   This is hardly surprising since the comoving number density of quasars at $M_\mathrm{B} \approx -28$\ and $z \gtrsim 1.5$\ is $\log \Phi(M_\mathrm{B}) \approx -8.15  $\ [Mpc$^{-3}$ mag$^{-1}$], and at $z \approx 0.5$\ and $M_\mathrm{B} \approx -23$\ is $\approx   -6.25$\ [Mpc$^{-3}$ mag$^{-1}$].} If  the integration limit used to compute $N_\mathrm{lum}$\ from --27.9 is lowered down to $M_\mathrm{B} \approx $ --26 at high-$z$, and if the luminosity function is extrapolated to this absolute magnitude,  the number of quasars $N_\mathrm{lum} $\ becomes comparable to $N_\mathrm{B}$ at low $z$. Thus, we cannot exclude that a quasar reached \mbh \ $\sim 10^{9}$ \msol\ at $z \sim 2$, stopped accreting, and was then rejuvenated at recent cosmic times.

\subsubsection{From seed to dead supermassive black holes}

Recent deep observations of  faint quasars ($m_\mathrm{V} \sim 22$)  obtained with GTC indicate the presence of a slowly evolving quasar Population at $z \approx 2$, not dissimilar to the one observed at low-$z$\ as far as the frequency of  Pop. B sources is concerned \citep{sulenticetal14}. This result is consistent with the existence of a  population of sources that were evolving on timescales much shorter than the Hubble time, then (at a cosmic age of just $\approx$ 3 Gyr) and now, and with the relatively short lifetimes expected for quasars  \citep{kellyetal10}.  As mentioned earlier, the comoving density of the very luminous quasars is very low, since they appear at the high end of the luminosity function. Therefore, it may be that -- if we exclude the most extreme sources -- we are seeing a process  going on systematically over cosmic epoch with quasar formation and evolution occurring in a way not so much different to the one expected for a human population (looking back in time, we can still identify adults and young adults as we see them at present). 

\citet{mathur00} and, independently, \citet{sulenticetal00a} suggested that the local-Universe NLSy1 sources accreting at a high rate are reminiscent of the early quasars.  We are still far from detecting the first population of ``infant'' quasars without ``adults'' (a feat that may never become possible). Unfortunately,  at high $z$ it is still not possible to detect black holes of \mbh $\sim 10^{7}$ \msol, even if they are radiating at or slightly above \lledd, as there is a redshift-dependent cut-off in the detectable \lledd\ \citep{sulenticetal14}.  Our view of quasars is biased. At intermediate to high-$z$\ we are ``blinded'' by the most luminous quasars in the Universe (such as the ones revealed by the Hamburg ESO survey, \citealt{wisotzkietal00}), objects whose luminosity may also be enhanced by anisotropic emission \citep{urryetal91,dipompeoetal14}. However,  these sources are the ones expected to exert a feedback effect significant enough to affect the host galaxy.  

Can we say that the local quasars are repeating the evolutionary patterns of these most luminous quasars? In the context of quasars, feedback effects significant enough to lead to the expulsion of matter on galactic scale may be possible only if the radiative and kinetic powers are extremely high \citep[e.g.,][see also the discussions in \citealt{marzianietal16,marzianietal16a}]{wagneretal13,kingpounds15}. A significant feedback effect of present-day luminous AGNs is proven only  in their circumnuclear regions, as shown in the very detailed study of the outflow in NGC 5548 \citep{kaastraetal14}, and energies involved in dispersing the gas outside of the bulge and in a host disk may be beyond the reach of present-day quasars. Why today we do not see them as the more luminous? The answer is related to the lack of available gas to get high accretion rates, and to the dramatic decrease of the expected merger rate with cosmic epoch \citep[e.g.,][]{cavalierevittorini00,hopkinsetal06}.} In keeping with a human population analogy, we can say that we see a population of young adults and adults which are similar, one close and one at a larger distance. At the larger distance we see a relatively rare population of quasars that are now extinguished. They grew so large that they destroyed their habitat and eventually starved to death.

  The cladogram we find shows that if evolution there is, it is very probably along the increasing \mbh, with a strong support from the physics and the demography of observed quasars. The difficulty is that the ontogeny describes the evolution of a single quasar. The demonstration in this section shows that a given quasar can only increase \mbh\ and in a relatively short time. If it appears as a Pop. A member, then it will likely evolve toward the Pop. B population. But the cladogram is not supposed to show this; rather, it depicts the phylogeny: if the rooting corresponds to the smaller \mbh, then the Pop. A quasars are closer to the most primitive quasars (as far as the diversity in our sample can tell of the diversity of the true population of quasars). This implies several things. First, a Pop. A quasars might not necessarily evolve toward Pop. B, there can be kind of dead-ends if for instance there is no more matter to accrete. These paths could be represented by (some of) the many branches of the trees. Second, and this is the main difference with ontogeny, there are many possible paths from the most ancestral groups. In other words, observing a  ``young adult'' does not tell {\em exactly} how the ``adult" will look like, even if physically, we can expect that a Pop. A member will have the typical Pop. B appearance. Finally, even if this possibility is probably not imaginable for local quasars, there is no element, from our study, to argue that a quasar cannot appear as such with a very high \mbh\ and high luminosity. 
 
 This is the limitation of rooting the tree with only one parameter,  {since quasar appearance cannot be reduced to the mass of their black hole}. However, the phylogeny of local quasars as revealed on our trees closely matches the ontogeny of their central black hole. This fact is supported, as mentioned at the end of the previous paragraph, by a physical motivation that emerges from the decrease of gas fraction and merger rate with increasing cosmic epoch. %

\begin{figure}[h!]
\begin{center}
\includegraphics[width=10cm]{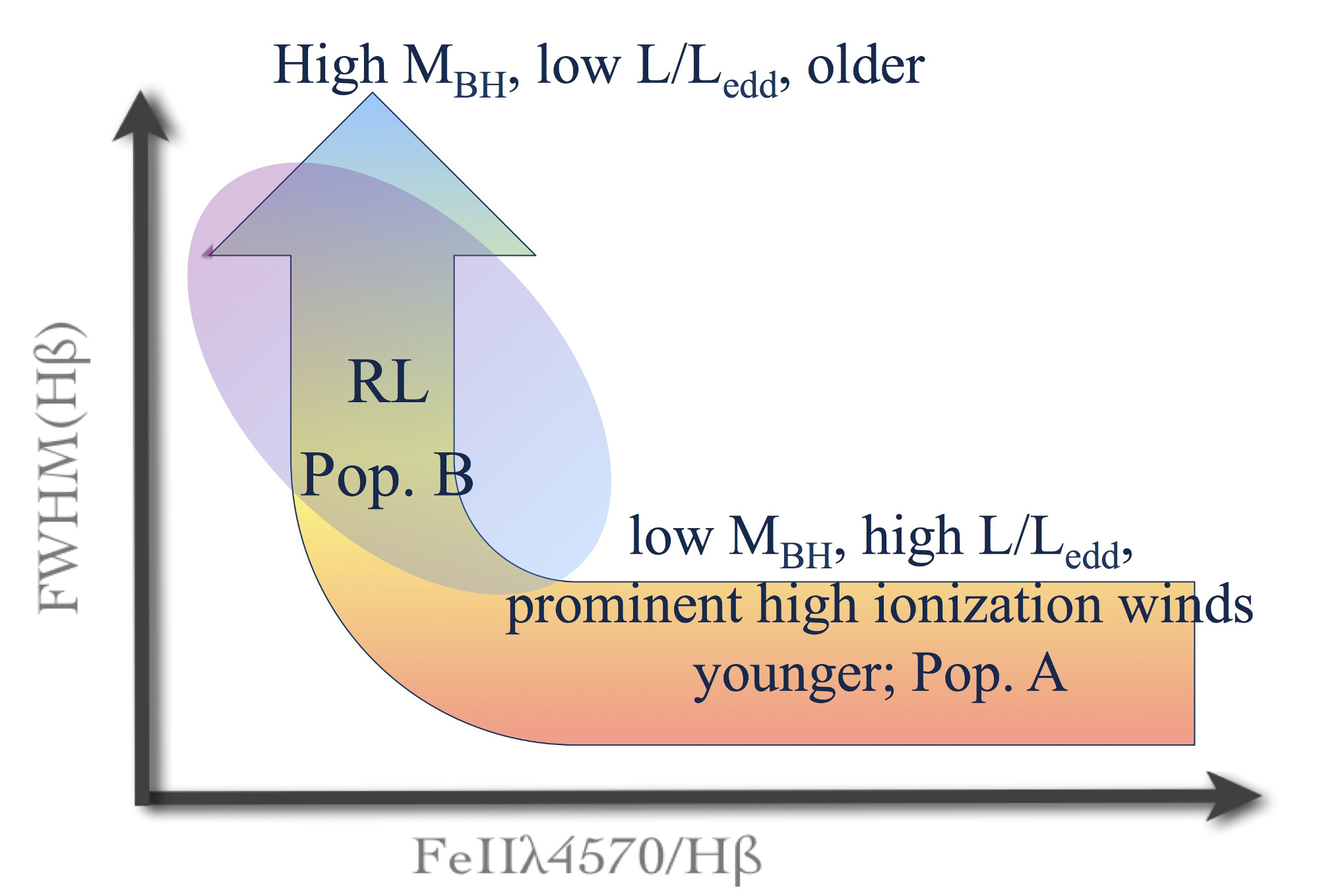}
\end{center}
\caption{A possible evolutionary scheme for RQ quasars. As \mbh\ increases, sources may move from the location of Population A sources radiating at relatively high Eddington ratio and with much evidence suggesting the presence of a radiation driven wind \citep{richardsetal11,sulenticetal07}, to Population B sources  in the optical plane of the 4DE1 parameter space.    } \label{fig:e1int} 
\end{figure}

\subsection{Radio-loud and radio-quiet quasars}
\label{rqrl}

Our cladogram reveals a \mbh\ threshold beyond which radio loudness could be triggered (\S\  \ref{threshold}). This is different from the threshold in the definition of RL quasars for which \citet{zamfiretal08} suggested a value that exceeds by an order of magnitude the canonical limit introduced by \citet{kellermannetal89}: $\log$ \rk\ $\gtrsim$ 1.8, and/or $\log P \gtrsim 31.6$ \ergss\ Hz$^{-1}$. RL sources in our sample are very powerful non thermal emitters and strongly deviate from the correlation between far-infrared (FIR) and radio expected for non-active galaxies \citep{bonzinietal15,padovani16}. The threshold we find corresponds to the occurrence of RL among the most massive objects, as has been already proposed by several workers \citep[e.g.,][and references therein]{chiabergemarconi11}, although this notion is   apparently challenged by the discovery of the so-called RL narrow-line Seyfert 1s \citep[e.g., ][]{komossaetal06,bertonetal16}.

At the same time, we cannot forget that in the optical plane of E1 we see sources that are both RQ and RL in the same area, in the radiatively-efficient domain associated with a thin accretion disk. Some features such as a strong redward asymmetry in the \hb\ profile \citep{marzianietal03b}, are observed in RQ and RL sources alike. RL and RQ Pop. B occupy also the same range in \mbh\ and \lledd, with modest \lledd, as shown in  samples larger than the one of the present paper \citep{sikoraetal07}.  Therefore, assuming that there is a \mbh\ threshold that makes it possible for an AGN to become RL, this condition is  necessary but not sufficient.  The threshold idea is appealing because it allows for the solution of the MCG 06-30-15 paradox: a maximally rotating black hole inferred from the Fe K$\alpha$ profile \citep{iwasawaetal96,sulenticetal98b} in a radio quiet quasar (actually an anonymous Seyfert 1 with an undistinguished spectrum, \citealt{sulenticetal98}).  The spin condition is therefore not a  sufficient condition  to guarantee radio loudness. 

In the context of the Blandford-Znajek mechanism for relativistic jet creation  \citep{blandfordznajek77}, the jet power is related to \mbh, the spin angular momentum $J$, and the magnetic field by:

\begin{equation}
P_{\nu}  = 5 \cdot 10^{10} \left(\frac{J}{J_\mathrm{max}} \right)^{2} \left(\frac{M_\mathrm{BH}}{1 M_{\odot}} \right)^{2}
\left(\frac{B}{\mathrm{1 G}} \right)^{2} \, {\rm erg \, s^{-1} Hz^{-1}},
\end{equation}
where $B$\ is the magnetic field and $J_\mathrm{max} =  GM^{2}/c$.  This equation indicates that even for a maximally rotating BH, and large mass ($M_\mathrm{BH} \sim 10^{9}$ \msol), a strong magnetic field is needed ($\sim 10^{2} - 10^{4}$ G). 

Recent theoretical work and numerical simulations indicate that jet collimation is occurring in a magnetically-arrested accretion disk (MAD) regime \citep{punsly15} where magnetic flux is maximized near the black hole, after being dragged from the outer disk toward the center.  A modest field strength expected to be present in the nuclei of galaxies ($\sim 1 \mu$ G) may yield a field strength close to the BH high enough to collimate the jet \citep{narayanetal03}.  However, whether a MAD will actually develop, depends on how the magnetic fields associated with the disk behaves as it is dragged toward the central black hole, a still debated issue. It is not understood under which conditions  the disk may be able to maintain its magnetization \citep[e.g.,][]{blandfordpayne82,riolsetal16}.  

Compact steep spectrum  and Giga-Hertz peaked sources \citep[][]{odea98} which often show evidence of high \lledd\ from optical and UV properties \citep{wu09b} are accounted for if MAD conditions and  jet collimation   also occur in an advection-dominated accretion flow (ADAF) context \citep[e.g.,][and references therein]{czernyyou16}  at high accretion rate i.e., for a geometrically thick, optically thick disk. A typical   case may be  3C 57, which hosts a very massive BH (\mbh $\sim 10^{9}$\ \msol, \citealt{sulenticetal15}), is very powerful at present and also shows evidence of relic radio emission, probably  due to a past activity cycle \citep{punslyetal16a}. Rejuvenated sources may be only a minority of cases, and hence account for the fact the radio-bright population is mostly due  to massive black holes radiating at modest  \lledd, as it is the case of the RL sources in M215 and M85. 

Therefore, if the attention is restricted to the most powerful RL sources, jet collimation may be made possible by the concomitant occurrence of high mass, high $J$, strong and ordered magnetic field. A corollary is that the RQ sources in the same \lledd\ and \mbh\ domain should have low values of $J$\ or of $B$.  According to   recent models, a relatively low \lledd\ ($\lesssim 0.1$) may not be a necessary condition, although it is highly unlikely  to have high \lledd\ for very massive BHs at low-$z$.   

\section{Conclusion}

The  cladistic analysis presented a new view of the quasar correlation space based on the first eigenvector of \citet{borosongreen92} and of \citet{sulenticetal00a}. In particular, it has been possible: (1) to identify the distinction between Pop. A and B, in a sequence of increasing \mbh; (2) to isolate RL sources among the most massive (more evolved) sources; (3) to separate CD and LD as monophyletic groups having the same progenitor. We infer, from these and previous results, that  Pop. B sources may indeed  be seen as   evolved Pop. A. We tentatively suggest that the radio quiet / radio loud dichotomy is influenced by  differences  in spin as well as by a different magnetization of the accretion disk.

\begin{table}[!t]
\begin{center}
\vbox{\footnotesize\tabcolsep=3pt
\parbox[c]{184mm}{\baselineskip=10pt
{Table 1.}{
\sc Main Trends Along the 4DE1 Sequence}}
\begin{tabular}{lccl}
\hline\hline
Parameter &  Population A 
& Population B &  References \\
\hline
FWHM(H$\beta_\mathrm{BC}$) & 800 -- 4000 km s$^{-1}$\ & 4000 -- 10000 km s$^{-1}$\ & 1, 2, 3, 4\\
$R_\mathrm{Fe}$\                         & 0.7              & 0.3                & 1, 2 \\
$c(\frac{1}{2})$ C{\sc iv}$\lambda$1549$_\mathrm{BC}$\     & --800 km s$^{-1}$\             & $-250$/+70 (RQ/RL)&5, 6, 7, 8\\
$\Gamma_{\mathrm S}$                 & often large   ($>$ 2)       & rarely large ($\approx$ 2) & 2, 9, 4, 10\\
\\
W(H$\beta_\mathrm{BC}$)                      & $\sim$ 80 \AA\   & $\sim$ 100 \AA\     & 2\\
H$\beta_\mathrm{BC}$\ profile shape          & Lorentzian       & double Gaussian    &  11, 12, 13\\
$c(\frac{1}{2}$) H$\beta_\mathrm{BC}$\       & $\sim$ zero       & +500 km s$^{-1}$\         & 13\\
{Si}{\sc iii} / {C}{\sc iii}] & 0.4              & 0.2                & 14, 15,16 \\
FWHMC{\sc iv}$\lambda$1549$_\mathrm{BC}$\                   & (2--6) $ \cdot 10^3$\ km s$^{-1}$\ & (2--10)  $\cdot 10^3$ km s$^{-1}$\ & 5, 17 \\
W(C{\sc iv}$\lambda$1549$_\mathrm{BC}$)                 & few \AA\  -- $\approx 60$ \AA\           & $\sim$ 100 \AA\ & 4, 6, 7\\
AI(C{\sc iv}$\lambda$1549$_\mathrm{BC}$)                & --0.1             & 0.05 & 5\\
W([O{\sc iii}$\lambda$5007)                &  1 -- 20\             & 20 -- 80 & 1, 18, 19 \\
v$_\mathrm{r}$([O{\sc iii}$\lambda$5007)                &     negative / 0         & $\sim 0 $ \kms\ & 4, 18, 19, 20 \\
\\
FIR color $\alpha(60,25)$   &  0 -- --1 & --1 -- --2 & 21\\ 
\\
X-ray variability         & extreme/rapid common & less common & 22, 23\\
optical variability       & possible             & more frequent/higher amplitude & 24\\
probability radio loud    & $\approx$ 3--4\%              & $\approx$ 0.25 \%              & 4,  25 \\
BALs                      &  extreme BALs &  less extreme BALs                   & 36,37 \\
\\
$ \log$ density$^{1}$          & $\gtrsim$11            & $\gtrsim$9.5        & 14, 28\\
$ \log U ^{1}$              &  --2.0/--1.5            &  --1.0/--0.5            & 14, 28\\
$\log M_\mathrm{BH}$ [\msol]          & 6.5 -- 8.5       & 8.0 -- 9,5          &7, 8, 29\\
$ L/L_\mathrm{Edd}$            & $\approx $ 0.2 -- 1.0        & $\sim $ 0.01 -- $\approx $ 0.2        & 1, 4,  7,  29, 30, 31\\
 \hline
\end{tabular}

  } 
\end{center}
\vskip-4mm

\medskip

\small{
1: \citealt{borosongreen92}; 
2:  \citealt{sulenticetal00a}; 
3: \citealt{collinetal06}; 
4: \citealt{shenho14}; 
5: \citealt{sulenticetal07};  
6:  \citealt{baskinlaor05b}; 
7: \citealt{richardsetal11}; 
8: \citealt{sulenticetal16}; 
9:  \citealt{wangetal96}:
10: \citealt{benschetal15}; 
11:  \citealt{veroncettyetal01};  
12:  \citealt{sulenticetal02};  
13:  \citealt{marzianietal03b}; 
 14:  \citealt{marzianietal01}; 
 15: \citealt{willsetal99}; 
 16: \citealt{bachevetal04}; 
 17: \citealt{coatmanetal16}; 
18:\citealt{zhangetal11}; 
19: \citealt{marzianietal16}; 
: \citealt{zamanovetal02}; 
21: \citealt{wangetal06};
22:  \citealt{turneretal99}; 
23:  \citealt{grupeetal01}; 
24:  \citealt{giveonetal99}; 
25: \citealt{zamfiretal08}; 
26:  \citealt{reichardetal03};
27: \citealt{sulenticetal06}; 
28: \citealt{negreteetal12}; 
29: \citealt{boroson02}; 
30:  \citealt{petersonetal04}; 
31: \citealt{kuraszkiewiczetal00}.
\medskip

$^{1}$ {Refers to a dense, low ionization emitting region that is apparently varying  along the E1 sequence, and is associated with the reverberation response of \hb\ (Negrete et al. 2013).}

} \end{table}

\section*{Conflict of Interest Statement}

The authors declare that the research was conducted in the absence of any commercial or financial relationships that could be construed as a potential conflict of interest.


\section*{Funding}

DD  acknowledges support from grants PAPIIT108716, UNAM, and CONACyT221398.

\section*{Acknowledgments}


\bibliographystyle{frontiersinSCNS_ENG_HUMS} 

\end{document}